\documentclass[aps,prl,reprint,superscriptaddress,showpacs]{revtex4-1}

\usepackage{dcolumn,amsmath,amssymb}
\usepackage{graphicx, mathtools}
\usepackage{epstopdf}
\usepackage{siunitx, bm, times, color}
\usepackage{csquotes}

\begin{document}

\title{Revealing three-dimensional structure of individual colloidal crystal grain by coherent x-ray diffractive imaging}

\author{A.G. Shabalin}
\affiliation{Deutsches Elektronen-Synchrotron DESY, Notkestr. 85, D-22607 Hamburg, Germany}
\affiliation{A.V. Shubnikov Institute of Crystallography RAS, Leninskii pr. 59, 119333 Moscow, Russia}
\author{J.-M. Meijer}
\affiliation{Van 't Hoff laboratory for Physical and Colloid chemistry, Debye Institute for Nanomaterials Science, Utrecht University, Padualaan 8, 3508 TB Utrecht, The Netherlands}
\affiliation{Present address: Division of Physical Chemistry, Department of Chemistry, Lund University, 22100 Lund, Sweden}
\author{R. Dronyak}
\affiliation{Deutsches Elektronen-Synchrotron DESY, Notkestr. 85, D-22607 Hamburg, Germany}
\author{O.M. Yefanov}
\affiliation{Deutsches Elektronen-Synchrotron DESY, Notkestr. 85, D-22607 Hamburg, Germany}
\affiliation{Present address: Center for Free-Electron Laser Science, DESY, Notkestr. 85, 22607 Hamburg, Germany}
\author{A. Singer}
\affiliation{Deutsches Elektronen-Synchrotron DESY, Notkestr. 85, D-22607 Hamburg, Germany}
\affiliation{Present address: University of California - San Diego, 92093 La Jolla, USA}
\author{R.P. Kurta}
\affiliation{Deutsches Elektronen-Synchrotron DESY, Notkestr. 85, D-22607 Hamburg, Germany}
\affiliation{Present address: European XFEL GmbH, Albert-Einstein-Ring 19, D-22761 Hamburg, Germany}
\author{U. Lorenz}
\affiliation{Deutsches Elektronen-Synchrotron DESY, Notkestr. 85, D-22607 Hamburg, Germany}
\author{O.Y. Gorobtsov}
\affiliation{Deutsches Elektronen-Synchrotron DESY, Notkestr. 85, D-22607 Hamburg, Germany}
\affiliation{NRC «Kurchatov Institute», pl. Akademika Kurchatova, Moscow, 123182, Russia}
\author{D. Dzhigaev}
\affiliation{Deutsches Elektronen-Synchrotron DESY, Notkestr. 85, D-22607 Hamburg, Germany}
\affiliation{National Research Nuclear University MEPhI (Moscow Engineering Physics Institute), Kashirskoe shosse 31, 115409 Moscow, Russia}
\author{S. Kalbfleisch}
\affiliation{Institute for X-Ray Physics, Friedrich-Hund-Platz 1, 37077 G\"{o}ttingen, Germany}
\author{J. Gulden}
\affiliation{Deutsches Elektronen-Synchrotron DESY, Notkestr. 85, D-22607 Hamburg, Germany}
\author{A.V. Zozulya}
\affiliation{Deutsches Elektronen-Synchrotron DESY, Notkestr. 85, D-22607 Hamburg, Germany}
\author{M. Sprung}
\affiliation{Deutsches Elektronen-Synchrotron DESY, Notkestr. 85, D-22607 Hamburg, Germany}
\author{A.V. Petukhov}
\affiliation{Van 't Hoff laboratory for Physical and Colloid chemistry, Debye Institute for Nanomaterials Science, Utrecht University, Padualaan 8, 3508 TB Utrecht, The Netherlands}
\affiliation{Laboratory of Physical Chemistry, Department of Chemical Engineering and Chemistry, Eindhoven University of Technology,  P.O. Box 513, 5600 MB Eindhoven,The Netherlands}
\author{I.A. Vartanyants}
\altaffiliation[Corresponding author: ]{ivan.vartaniants@desy.de}
\affiliation{Deutsches Elektronen-Synchrotron DESY, Notkestr. 85, D-22607 Hamburg, Germany}
\affiliation{National Research Nuclear University MEPhI (Moscow Engineering Physics Institute), Kashirskoe shosse 31, 115409 Moscow, Russia}

\date{\today}

\begin{abstract}

We present results of a coherent x-ray diffractive imaging experiment performed on a single colloidal crystal grain.
The full three-dimensional (3D) reciprocal space map measured by an azimuthal rotational scan contained several orders of Bragg reflections together with the coherent interference signal between them.
Applying the iterative phase retrieval approach, the 3D structure of the crystal grain was reconstructed and positions of individual colloidal particles were resolved.
As a result, an exact stacking sequence of hexagonal close-packed layers including planar and linear defects were identified.

\end{abstract}


\maketitle

%

Colloidal crystals nowadays are actively exploited as an important model system to study nucleation phenomena in freezing, melting and solid-solid phase transitions~\cite{qi2015nonclassical,wang2015direct,statt2015finite,peng2015two}, jamming and glass formation~\cite{degiuli2014force,jacob2015convective}.
In addition, colloidal crystals are attractive for multiple applications since they can be used as large-scale templates to fabricate novel materials with unique optical properties such as the full photonic bandgap, 'slow' photons and negative refraction, as well as materials for application in catalysis, biomaterials and sensorics~\cite{kim2011self,kuzyk2012dna,henry2011crystallization,zhang_lu2013general}.
Colloidal crystals provide a low cost large-scale alternative to lithographic techniques, which are very effective for producing high-quality materials with a desired structure, but are limited in building up a truly three-dimensional (3D) structures and bring high production costs~\cite{painter1999two}.
Recently, significant progress is achieved in engineering materials with tunable periodic structure on the mesoscale by functionalizing colloids with DNA \cite{zhang2013general}, applying external fields \cite{yethiraj2003colloidal,pal2015tuning} or varying the particle shape \cite{glotzer2007anisotropy,petukhov2015particle}.

Understanding the real structure of colloidal crystals and disorder of different types is an important aspect from both fundamental and practical points of view. Even at equilibrium colloidal crystals can have a finite density of defects, which can be anomalously large for certain colloidal lattices~\cite{smallenburg2012vacancy}. The opposite can also happen as defects can play a decisive role in the choice of the crystal structure~\cite{mahynski2014stabilizing,hilhorst2010slanted}. For applications such as photonic crystals most of growth-induced defects can deteriorate their optical properties. On the other hand, controlled incorporation of certain defects can be desirable to enhance the functionality such as creating waveguides \cite{johnson2001photonic,hilhorst2009double}, trapping photons~\cite{vlasov2001chip, painter1999two} and developing optical chips~\cite{soljavcic2003nonlinear}. In these studies monitoring an internal 3D structure of colloidal crystals including defects in real time is an important aspect, which remains a challenge \cite{Sulyanova2015}.

Among widely used techniques of the colloidal crystals structure investigation are optical microscopy~\cite{dinsmore2001three,schall2004visualization} and confocal laser scanning microscopy~\cite{schall2009laser}.
However, the range of applications of these methods is strongly reduced by the limited resolution (at best about \SI{250}{\nano\meter}) and the need of a careful refractive index matching, which is not always possible. Furthermore, some of the materials are opaque for visible light which complicates imaging of their internal structure.
Electron microscopy (EM) can provide high-quality images of the material surface with an exceptional resolution~\cite{jiang1999single}, but fails to probe the bulk, because of the short penetration depth of electrons.
In addition, imaging in EM typically involves elaborate and destructive sample preparation~\cite{ye2002self}, such as drying or coating.
In this respect, high-resolution x-ray microscopy~\cite{bosak2010high, hilhorst2012three, byelov2013situ, van2011scanning}, small-angle x-ray scattering (SAXS)~\cite{sirota1989complete} and small-angle neutron scattering (SANS)~\cite{chen1994rheological} represent complementary methods offering the advantage of high penetration depth for nondestructive studies of colloidal systems.

Here we present results of coherent x-ray diffractive imaging (CXDI) approach \cite{miao1999extending, robinson2001reconstruction} (see also for review \cite{vartanyants2015coherent}) which allows to visualize the internal 3D structure of an individual colloidal crystal grain with high resolution.

The original idea was proposed more than half a century ago in a seminal work of Sayre~\cite{sayre1952squaring}, where he suggested to phase crystallographic data by measuring information between the Bragg peaks in reciprocal space \footnote[1]{Demonstration of this approach to reveal the structure of a single protein was given in a recent work performed at x-ray free-electron lasers~\cite{ayyer2016macromolecular}}.
In the context of CXDI technique this concept can be applied for three-dimensional imaging of the internal structure of crystalline samples and, in principle, can provide information about the positions of individual scatterers in the crystalline structure~\cite{vartanyants2015coherent}.
We have demonstrated in simulations that such approach can be applied to atomic resolution imaging of nanocrystals illuminated with high energy coherent x-rays by mapping several Bragg peaks on the detector \cite{gulden2011imaging} (similar to high energy electron scattering ~\cite{zuo2003atomic, dronyak2010electron, chen2013three}).
However, lack of coherent flux as well as low scattering efficiency at these energies are strong limiting factors to reach these goals presently.

At the same time, colloidal crystals with the unit cell on the order of few hundred nanometers are an ideal objects for developing these high resolution imaging methods.
First experimental demonstration of applying two-dimensional (2D) CXDI method for visualization of a single stacking fault in a thin colloidal crystalline film was presented in Ref.~\cite{gulden2010coherent}.
Due to experimental challenges attempts to generalize this approach to 3D were not successful up to now~\cite{gulden2012three}.
Here we present first successful results of a detailed reconstruction, which visualize three-dimensional positions of individual particles in a single colloidal crystal grain.

\begin{figure}
\centerline{\includegraphics[angle=0, width=8.6cm]{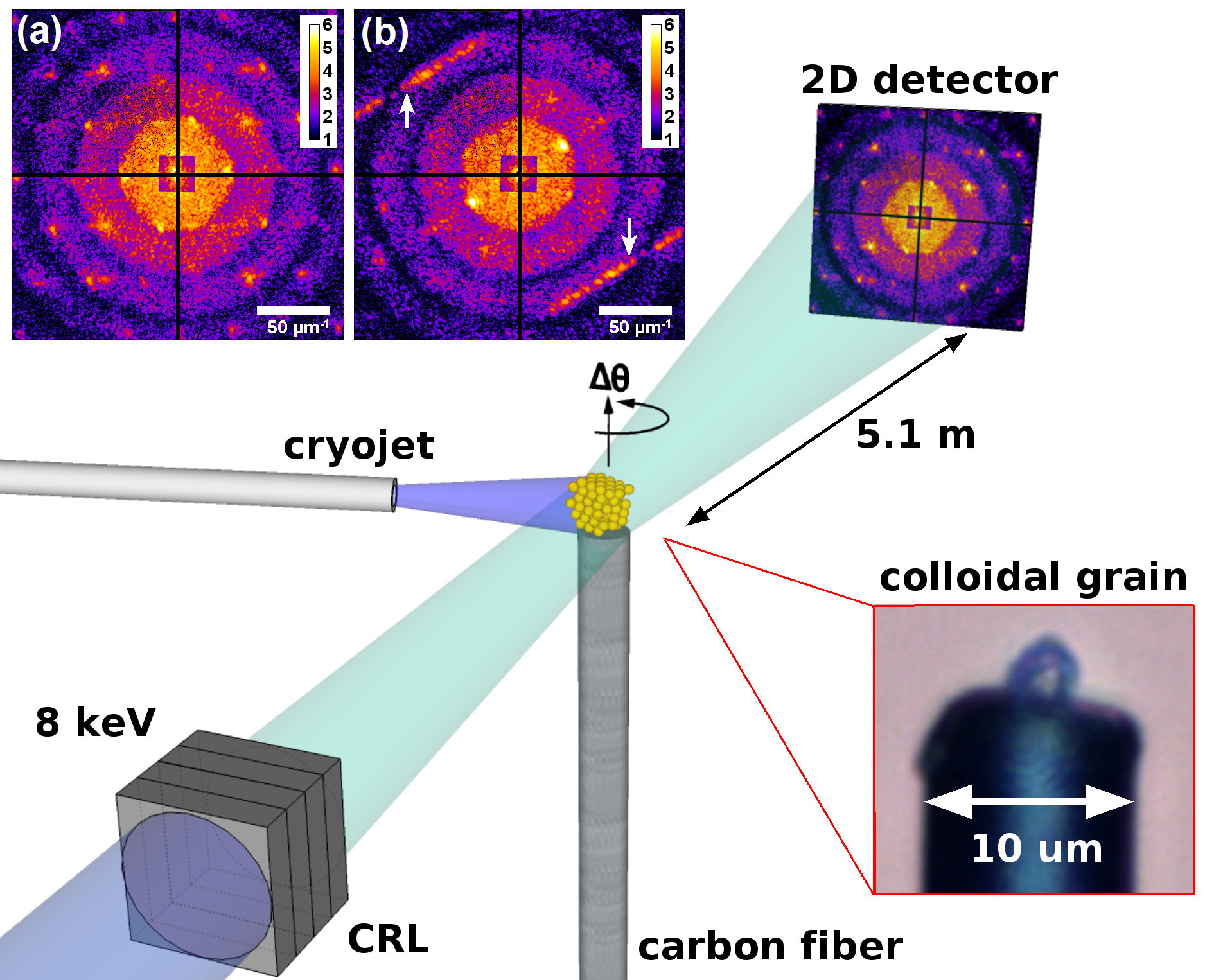}}
	\caption{Schematic layout of the experimental setup.
		Compound refractive lenses (CRL) focus the coherent x-ray beam at the crystal grain mounted on a top of the carbon fiber (light microscope image is shown in lower right inset).
		The sample was constantly cooled by a flow of nitrogen using a cryojet. The diffraction data were recorded by 2D detector positioned in the far-field.
		(a, b) Typical diffraction patterns in logarithmic scale measured at relative angular positions $\Delta\theta=\SI{61}{\degree}$ (a) and $\Delta\theta=\SI{165}{\degree}$ (b).
		Two elongated rods visible in (b) (indicated by arrows) originate from the presence of a planar defect in the crystal grain.}
	\label{fig:Sketch}
\end{figure}

The experiment was performed at the Coherence Beamline P10 at PETRA III in Hamburg (see for experimental details also Appendix I).
A monochromatic coherent x-ray beam of \SI{8}{\kilo\electronvolt} photon energy was focused at the sample at \SI{87.7}{\meter} from the undulator source using the transfocator optics~\cite{zozulya2012microfocusing} based on compound refractive lenses (CRLs) positioned at \SI{2.2}{\meter} distance upstream from the sample (see Figure~\ref{fig:Sketch}).
A set of horizontal and vertical guard slits $\num{75} \times \num{75} $ $\SI{}{\micro\meter}^2$ in size, located at  \SI{1.5}{\meter} distance in front of the CRLs, was used to select a coherent portion of the beam.
The size of the focal spot at the sample position was \SI{5.5}{\micro\meter} by \SI{3.2}{\micro\meter} full width at half maximum (FWHM) in horizontal and vertical directions, with the total intensity about $\num{10}^{11}$ photons per second ~\cite{zozulya2012microfocusing}.
To inhibit the radiation damage the sample was cryo-cooled with the flow of nitrogen (about \SI{100}{\kelvin}).
The sample holder was mounted on a goniometer, which allows azimuthal rotation around the vertical axis.
The diffraction data were recorded using a photon-counting pixel detector  MAXIPIX positioned in transmission geometry at  \SI{5.1}{\meter} distance downstream from the sample.
In order to reduce air scattering an evacuated tube was inserted between the sample and the detector covering the major part of the optical path.
The total number of pixels of the detector was $\num{516} \times \num{516}$ and a pixel size was $\num{55} \times \num{55}$ $\si{\micro\meter}^2$. 
For the selected photon energy and sample-to-detector distance resolution in reciprocal space was $\SI{0.437}{\micro\meter^{-1}}$ per pixel and allowed to have four times sampling rate per speckle.

Samples were prepared from dried sediments of colloidal crystals that showed characteristic optical Bragg reflections~\cite{gulden2012three}.
The specimen studied in this work consisted of sterically stabilized silica spheres with a diameter of \SI{230}{\nano\meter} dried from cyclohexane over several months.
Small grains were obtained by mechanically crushing a piece of the ordered sediment.
Individual grains were picked up using a micromanipulator and connected to the tip of a \SI{10}{\micro\meter} thick carbon fiber, which was glued to a glass holder beforehand.
The colloidal crystal grain used in this study was imaged with a light microscope (see inset in Figure~\ref{fig:Sketch} and Appendix I) and determined to have dimensions of about  $\num{2} \times \num{3} \times \num{4}$ $\SI{}{\micro\meter}^3$.

The full dataset consisted of rotation series of \num{360} diffraction patterns with \SI{0.5}{\degree} angular increment covering the entire reciprocal space.
To avoid oversaturation of the detector a series of \num{50} images with \SI{0.02}{\second} of exposure time were collected and summed up for each azimuthal position.
In Figure~\ref{fig:Sketch}(\textit{a},\textit{b}) two examples of obtained scattering patterns are shown.
They contain several Bragg peaks surrounded by the interference speckles and diffuse scattering in between.
The detector size allowed to record reflections up to the forth order measured simultaneously.
The visibility $V=(I_{max}-I_{min})/(I_{max}+I_{min})$,
%
%
which is a commonly used parameter to describe the contrast in the coherent interference pattern~\cite{goodman2005introduction}, was estimated to be from \SIrange[]{75}{80}{\percent}.
It is worth to mention, that although the intensities between Bragg peaks are rather weak in comparison with the ones in the vicinity of Bragg peaks, this part of the recorded signal is highly important for the reconstruction.
It encodes information about the relative phases of different reflections and allows to resolve positions of individual scatterers in the unit cell.
The concentric rings observed in diffraction patterns represent the form factor of a single colloidal sphere and the number of these rings in the diffraction pattern
gives us an estimate of four pixels resolution per colloidal particle diameter in direct space.
%
\begin{figure}
		\includegraphics[angle=0, width=8.0cm]{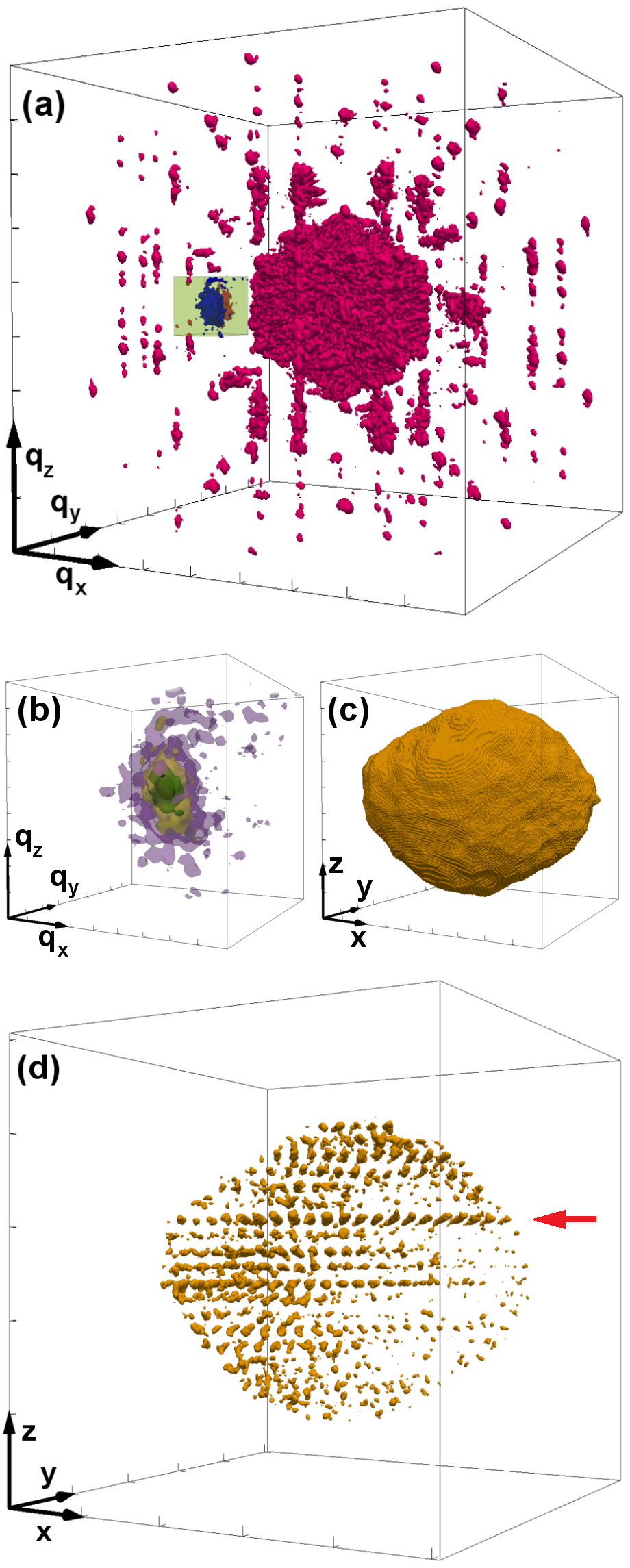}
		\caption{
			(a) 3D scattered intensity in reciprocal space, represented by volume rendering.
			(b) Reciprocal space map in the vicinity of the Bragg peak shown by a green box in (\textit{a}).
			(c) Averaged reconstruction of 3D shape of the grain in real space obtained from (\textit{b}).
			(d) Reconstructed 3D electron density distribution in real space. The red arrow marks the crystalline plane shown as 2D slice in Figure~\label{fig:Structure}(c).
			Length of the coordinate arrows correspond to $\SI{50}{\micro\meter^{-1}}$ in (\textit{a}), $\SI{10}{\micro\meter^{-1}}$ in (\textit{b}), and  $\SI{1}{\micro\meter}$ in (\textit{c}, \textit{d}).}
			\label{fig:3D}
\end{figure}
%


The full 3D reciprocal space map, represented by volume rendering, is shown in Figure~\ref{fig:3D}(\textit{a}).
All \num{360} diffraction patterns collected during the rotational scan were used in the merging procedure.
A remarkable features observed in this map are a number of streaks which connect some of the Bragg peaks (see Figure~\ref{fig:3D}(\textit{a}) and diffraction pattern shown in Figure~\ref{fig:Sketch}(\textit{b})).
These streaks (Bragg rods) indicate the presence of plane defects in the crystalline lattice and the intensity modulations along them are directly related to the exact stacking sequence~\cite{gulden2012three, meijer2014double}.
In the full 3D reciprocal space dataset (see Figure~\ref{fig:3D}(\textit{a})), that gives rise to well-pronounced Bragg rods which connect some of the reciprocal lattice nodes.
Such rods are oriented perpendicular to the plane of the defect and therefore a specific sample orientation is required for the observation of the whole streak in a single diffraction pattern.

%

The 3D reciprocal space maps were inverted to real space images by using a phase retrieval algorithm \cite{fienup1982phase,marchesini2007}.
In order to obtain a good estimate for a tight support, which is crucial to enforce convergence of the reconstruction procedure, we analyzed the scattered intensity distribution in the vicinity of several Bragg reflections that were not affected by Bragg rods.
From each of them we selected a cubic volume of $\num{35}\times \num{35} \times \num{35}~\si{\micro\meter}^{-3}$ surrounding the Bragg peak (see Figure~\ref{fig:3D}(b)), and performed reconstruction for this cropped dataset.
The results were averaged over reconstructions of six different Bragg reflections, and the obtained shape function  (see Figure~\ref{fig:3D}(c)) was used as a tight support in the reconstruction of the full 3D reciprocal space dataset.
That allowed to avoid appearance of twin images and substantially facilitated the reconstruction procedure.
In the phase retrieval process, consisted of about \num{3000} iterations of Hybrid Input-Output (HIO) combined with Error Reduction (ER) algorithms, the support was several times updated by applying the Shrinkwrap method~\cite{fienup1982phase, marchesini2003x}.
The missing regions in the diffraction data were allowed to freely evolve with an additional constraint of applying an upper boundary for amplitudes. This suppresses them down to values obtained from the Fourier transform of the support with normalization to the measured amplitudes.

Reconstructed 3D electron density after applying a high band pass Gaussian filter is shown in Figure~\ref{fig:3D}(d).
It reveals periodic behavior which corresponds to positions of colloidal particles in the crystal grain.
In Figure~\ref{fig:Structure}(b) a slice through one of the packing planes marked by a red arrow in Figure~\ref{fig:3D}(d) is presented.
Remarkably, positions of individual colloidal spheres are clearly resolved and show perfect hexagonal symmetry.
We used the same slice to estimate an obtained resolution in real space for localization of individual particles in the crystal.
An average intensity peaks width in this image was estimated to be \SI{70}{\nano\meter} (FWHM).
Applying Rayleigh criterion for the minimum separation between two distinguished Gaussian peaks we determined the resolution to be on the order of \SI{80}{\nano\meter}.

Furthermore, our results allow to identify the stacking sequence of the colloidal crystal layers.
The two most common close-packed structures which occur in nature are the hexagonal close-packed (HCP) structure with a stacking period AB, and the face-centered cubic (FCC) with a layer stacking of ABC (see Fig.~\ref{fig:Structure}(a)).
In a colloidal crystal the free-energy difference between HCP and FCC is rather small~\cite{woodcock1997entropy}.
Therefore a random mixture of these two stacking types, the so-called random hexagonal close-packed (RHCP) structure, is often observed in colloidal crystals spontaneously self-assembled under gravity~\cite{petukhov2003bragg, dolbnya2005coexistence}.



To determine the stacking sequence (see also Appendix II) we analyzed the projection of the 3D density map along the [100] crystallographic direction of the hexagonal lattice defined in Ref.~\cite{meijer2014double} \emph{i.e.} along the x-axis in Figure~\ref{fig:3D}(d) .
To avoid error accumulation in the projection, each position of the intensity maxima in the 3D density profile was substituted by identical 3D Gaussian functions with the FWHM of \SI{70}{\nano\meter} and no density in between them.
Next, for the sake of convenient visual representation, the resulting density map was divided by a 2D profile of the averaged intensity distribution, as a flat field correction to the image.
In Figure~\ref{fig:Structure}(c) each maximum represents a column of spherical particles in a crystalline arrangement.
In the case of an ideal close packed structure, projection of a column would correspond to projection of a single sphere.
Non-uniform intensity distribution presented in Figure~\ref{fig:Structure}(c) can be attributed to imperfections in positioning of individual particles (see \cite{Supplement}).
Since the vector of lateral displacement belongs to the projection plane, the lateral position of each layer can be directly identified.
The first well pronounced layer at the top of the crystal was denoted as an A layer and the next layer below as a B layer.
Then, according to its lateral shift, the third layer appears to be a C layer.
Following this procedure \num{19} layers in the colloidal crystal were identified (see Figure~\ref{fig:Structure}(c)).
One of the possible sequences of elementary stacking order may be $\underbrace{ABC}\underbrace{ABCB}\underbrace{CBCA}\underbrace{CB}\underbrace{CBAB}\underbrace{AB}$ (from top to bottom) that reveals three packing structures: FCC, HCP, and double hexagonal close-packed (DHCP) structure.
The last one is usually described by $ABCB$ layer period (see Figure~\ref{fig:Structure}(a)).
Here we proposed one of the possible sequences of the elementary stacking order, however alternative decompositions are also possible.
Still, all of them will contain FCC, HCP, and DHCP blocks in different combinations.

In addition to regular structure revealed by our analysis, we have observed a number of linear defects in the projection of 3D colloidal crystal grain (see insets in Figure~\ref{fig:Structure}(c)).
In the inset I it is clearly seen that at this specific location the left part of the crystal sequence can be clearly identified as $B$ position, while the right part is rather $A$ position.
In the inset II different type of defect is shown, when several double maxima appear in the projection.
Our inspection showed that it can be attributed to boundary between two perfect hexagonal regions which are shifted by half period perpendicular to the projection direction.
We would like to point out that our reconstruction visualizes exact stacking sequence of layers and in-plane defects in colloidal crystal grain.
This is an important achievement which opens a way to further applications of CXDI method for non-destructive characterization of photonic crystals.

\begin{figure}
		\includegraphics[width=8.0cm]{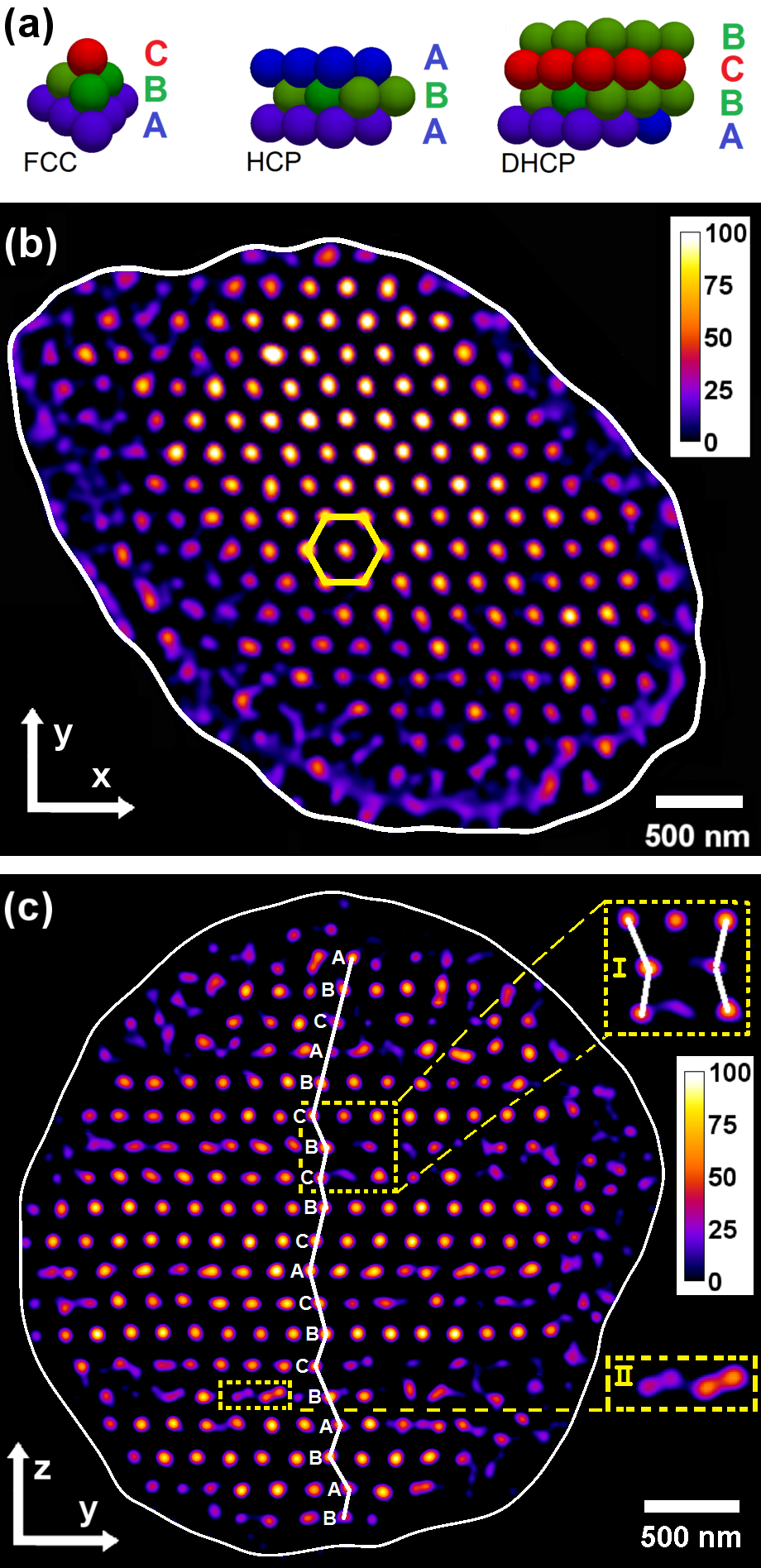}
		\caption{
		(a) Examples of regular close-packed structures with the corresponding layer sequences: face centered cubic (FCC), hexagonal close packed (HCP), and double hexagonal close
        packed (DHCP).
		(b) Slice through one of the packing planes.
        Well pronounced hexagonal symmetry is outlined by a yellow hexagon.
		(c) Projection of the density map on the [100] crystallographic direction of the hexagonal lattice.
		Each layer is marked by the corresponding letter, the insets (I-II) show examples of in-plane defects.
        }
		\label{fig:Structure}
\end{figure}
%

%
%



In a summary, non-destructive CXDI experiment on a single colloidal crystal grain was performed.
Full crystallographic data which included several Bragg reflections together with surrounding speckles and intensities between them was measured by a rotation series of 2D far-field diffraction patterns.
By applying the phase retrieval approach the obtained 3D dataset in reciprocal space was inverted into the electron density distribution in real space.
Finally, the grain shape and positions of individual colloidal particles, resolved in three dimensions, were reconstructed.
The crystalline structure of the sample was characterized in terms of close packing of hexagonal layers.
The determined stacking sequence revealed \num{19} layers with FCC, HCP, and DHCP structure blocks.
Our results allowed us to visualize with high resolution a number of in-plane defects present in the colloidal crystal grain including all the details up to its core.

Our results open up a breakthrough in applications of coherent x-ray diffraction for visualization the inner three-dimensional structure of different mesoscopic materials, such as photonic crystals.
The outcome of this work is of significant importance for further progress and developments of CXDI methods with an aim to resolve the three-dimensional structure of nanocrystals with atomic resolution ~\cite{gulden2011imaging}.
Our achievements pave the way to atomic resolution imaging of nanocrystals at the next generation diffraction limited synchrotron light sources \cite{hettel2014dlsr} that are expected to provide two orders of magnitude higher coherent flux.

\bigskip

\begin{acknowledgments}
This research was supported by BMBF Proposal 05K10CHG "Coherent Diffraction Imaging and Scattering of Ultrashort Coherent
Pulses with Matter" in the framework of the German-Russian collaboration Development and Use of Accelerator-Based Photon Sources and the Virtual Institute VH-VI-403 of the Helmholtz Association.
The use of GINIX setup operated by the University of G$\ddot{o}$ttingen is greatly acknowledged.
We acknowledge fruitful discussions and support of the project by E. Weckert, important comments and suggestions on the final stages of the manuscript preparation by
S. Lazarev and careful reading of the manuscript by D. Novikov.
\end{acknowledgments}

\begin{appendix}

\section{Appendix I. Experimental details}
\label{AppA}

\begin{figure}
	\includegraphics[angle=0, width=8.6cm]{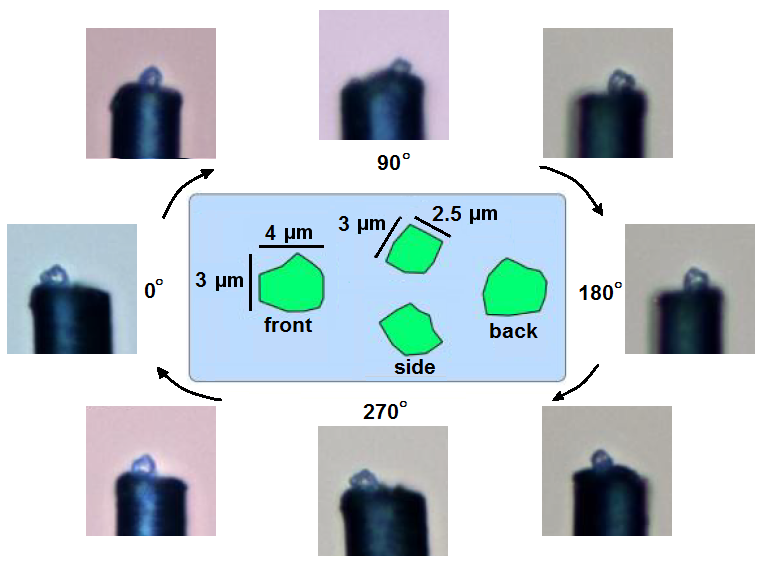}
	\caption{Light microscopy images of the colloidal crystal grain mounted on a carbon fiber tip. Perspectives from different azimuthal rotation angles are presented. The central panel shows estimates of the shape function in several projections. }
	\label{fig:LM}
\end{figure}

After the CXDI measurements the colloidal crystal grain was imaged with a light microscope.
A series of images taken from different perspectives covering \num{360} degrees in angular range is presented in  Figure~\ref{fig:LM}).
Estimates of the shape function in projection on different directions are shown in the central subset.
According to them the grain is determined to have dimensions of about  $\num{2} \times \num{3} \times \num{4}$ $\SI{}{\micro\meter}^3$.

To access the scattering signal close to the directly transmitted beam the semitransparent beamstop $5 \times 5 $ mm in size made of \SI{300}{\micro\meter} thick Si foil, was installed in front of the detector. In addition, a Ta disk of  \SI{0.5}{\milli\meter} in diameter was glued on the top of the foil to absorb the direct beam completely.
The semitransparent beamstop can be recognized as the shadowed area in the center of the diffraction images presented in
Figure~1(\textit{a},\textit{b}). 
Afterwards, the recorded signal in this region was multiplied by a mask, which was calculated according to the absorption length for the selected photon energy and the film thickness.
Such a simple rescaling, however, cannot be used near the beamstop edge (about \num{1}\--\num{2} pixels in width) along the perimeter of the shadowed area.
On one hand the borders of the foil could have variations of thickness in the consequence
of cutting procedure, on the other hand the detector pixels under this area could be shadowed partially.
To handle this the mask coefficient for each pixel in this region  was determined by normalization of the detected signal to the averaged value of intensity in the surrounding.
To improve accuracy the results of this normalization were averaged over all diffraction patterns collected during the measurements.

In addition to the rotation series several images were recorded with the empty-beam, i.e. when the sample was completely moved out. The average of these images was normalized to the expose time and subtracted as a background from each of the diffraction patterns in the dataset.


In Figure~2(a) of the main text the full 3D reciprocal space map, represented by volume rendering,  is shown.
The volume of $\num{223}~\times~\num{223}~\times~\num{223}~\si{\micro\meter}^{-3}$ is sampled by a regular grid of $\num{511}~\times~\num{511}~\times~\num{511}$ voxels, so the transversal size of the voxel corresponds to the pixel size of the diffraction pattern in reciprocal space.


\section{Appendix II. Stacking sequence determination}
\label{AppB}

\begin{figure*}
	\includegraphics[angle=0, width=17.2cm]{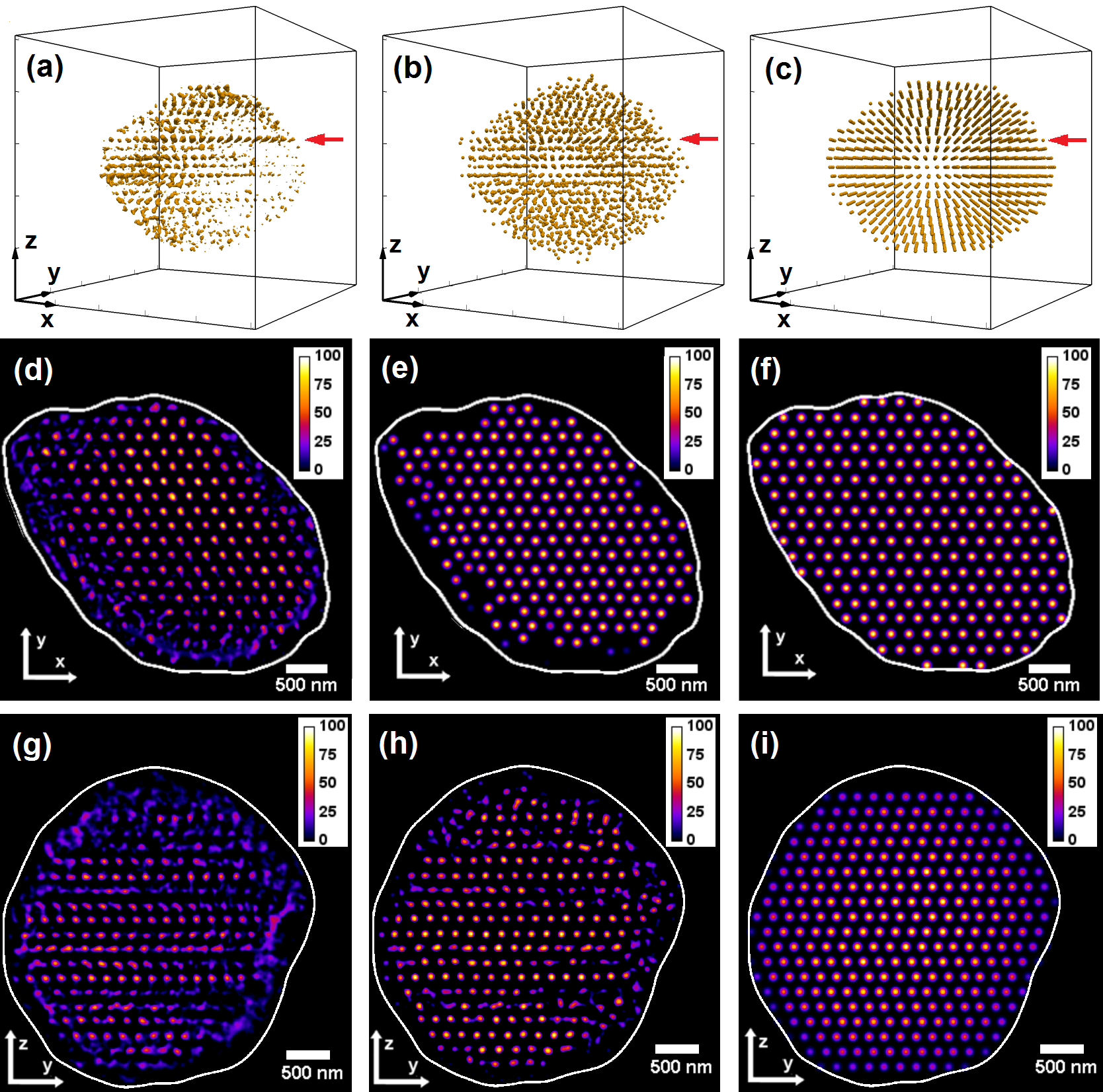}
	\caption{
		Results of reconstruction and the resolved positions of individual particles represented by balls placed at the positions of local maximums compared to modeled structure of a perfect crystal of the same shape and stacking sequence.
		(a-c) Iso-surfaces representing density distribution in real space.
		(d-f) Slice through a single hexagonal layer.
		(g-i) Projection on the [100] crystallographic direction.}
	\label{fig:Comparison}
\end{figure*}

From theory~\cite{krishna1981close} it is known, that highest averaged density in close-packed arrangements of equal spheres is achieved when they form plain hexagonal layers, which can occupy only three specific positions relative to each other, as shown in Figure~3(a) of the main text. 
Let the layer at the bottom be called A layer (colored with blue in schematic).
The layer above it can be placed in two types of triangular voids, one with the apex upwards and labeled B (green color), and the other with the apex downwards and labeled C (red color). Only one of these sites can be occupied, but not both.
If the second layer is B, then, similarly, the third hexagonal close-packed layer can occupy either A or C positions and so on.
Any sequence of letters, A, B and C with no two successive letters alike represents a possible manner of close-packing.
In such a 3D structure, each sphere is surrounded by and touches \num{12} other spheres.

The determined periodicity in the stacking sequence is a remarkable result because the DHCP structure has not been observed before in colloidal crystals of spheres.
A recent study~\cite{meijer2007plane} has indicated that during the sedimentary formation process of a colloidal crystal the layers nucleate sequentially.
The interactions between the layers are limited to the neighboring layers and results in the RHCP structure.
The formation of a DHCP structure could just be one of the many random coincidental realizations of the RHCP structure.
However, one cannot exclude the influence of the drying procedure as capillary forces of significant strength will act on the colloids during solvent evaporation~\cite{denkov1992mechanism} and might induce a structural reorganization.

Results of our reconstruction were also confirmed by an independent analysis of the reciprocal space data (see Ref.~\cite{meijer2014double}), based on a simplified structure model consisting of a finite number of equally-sized hexagonal close-packed layers.
In this model DHCP structure was also identified, however, results of the phase retrieval presented here provide much more detailed information about the structure of a colloidal grain.

To verify our reconstruction results we compared them with the model of a perfect crystal grain with the same shape composed of perfect hexagonal layers with the same stacking sequence.
In Figure~\ref{fig:Comparison}(a) originally reconstructed 3D electron density function is presented.
Next, in Figure~\ref{fig:Comparison}(b) it is compared with the results of replacement of the positions of local maxima by identical 3D Gaussian functions with \num{70} nm (FWHM).
In the last Figure~\ref{fig:Comparison}(c) we show a model crystal of the same shape composed of perfect hexagonal layers with the same stacking sequence as determined in our experiment.
In Figures~\ref{fig:Comparison}(d-f) slices through the corresponding 3D density maps are shown and in Figures~\ref{fig:Comparison}(g-i) the projections on the [100] crystallographic direction are presented.

Such a comparison proves high quality of the reconstruction. All the geometrical characteristics of the colloidal particles arrangement revealed by reconstruction, such as distances between individual particles, hexagonal symmetry within each layer and interlayer spacing are in excellent agreement with the model. A comparison of the reconstructed and modeled structures in the projection demonstrates correctness of our approach for the stacking sequence determination based on relative lateral displacements between neighboring layers.

%

\end{appendix}


\bibliography{./references}

\end{document}